\documentclass[proof]{WileyASNA-v1}

\articletype{Article Type}%

\received{2021}
\revised{2021}
\accepted{2021}

\begin{document}

\title{A small radio galaxy at $z=4.026$}

\author[1,2,3]{Krisztina \'E. Gab\'anyi*}

\author[2]{S\'andor Frey}

\author[4]{Tao An}

\author[5]{Hongmin Cao}

\author[6]{Zsolt Paragi}

\author[6,7]{Leonid I. Gurvits}

\author[4]{Yingkang Zhang}

\author[8]{Tullia Sbarrato}

\author[1,2]{M\'at\'e Krezinger}

\author[2]{Krisztina Perger}

\author[2]{Gy\"orgy Mez\H{o}}

\authormark{K. \'E. GAB\'ANYI \textsc{et al}}

\address[1]{\orgdiv{Department of Astronomy, Institute of Geography and Earth Sciences}, \orgname{ELTE E\"otv\"os Lor\'and University}, \orgaddress{\state{Budapest}, \country{Hungary}}}

\address[2]{\orgdiv{Konkoly Observatory}, \orgname{ELKH Research Centre for Astronomy and Earth Sciences}, \orgaddress{\state{Budapest}, \country{Hungary}}}

\address[3]{\orgdiv{ELKH-ELTE Extragalactic Research Group}, \orgname{ELTE E\"otv\"os Lor\'and University}, \orgaddress{\state{Budapest}, \country{Hungary}}}

\address[4]{\orgname{Shanghai Astronomical Observatory}, \orgaddress{\state{Shanghai}, \country{China}}}

\address[5]{\orgname{Shangqiu Normal University}, \orgaddress{\state{Shangqiu}, \country{China}}}

\address[6]{\orgname{Joint Institute for VLBI ERIC}, \orgaddress{\state{Dwingeloo}, \country{The Netherlands}}}

\address[7]{\orgdiv{Department of Astrodynamics and Space Missions}, \orgname{Delft University of Technology}, \orgaddress{\state{Delft}, \country{The Netherlands}}}

\address[8]{\orgdiv{Osservatorio Astronomico di Brera}, \orgname{INAF}, \orgaddress{\state{Merate}, \country{Italy}}}

\corres{*Krisztina \'E. Gab\'anyi \email{krisztina.g@gmail.com}}

\presentaddress{Department of Astronomy, E\"otv\"os Lor\'and University, P\'azm\'any P\'eter s\'et\'any 1/A, H-1117 Budapest, Hungary}

\abstract{Less than $200$ radio-loud active galactic nuclei (AGN) are known above redshift $4$. Around $40$ of them have been observed at milliarcsecond (mas) scale resolution with very long baseline interferometry (VLBI) technique. Some of them are unresolved, compact, relativistically beamed objects, blazars with jets pointing at small angles to the observer's line of sight. But there are also objects with no sign of relativistic beaming possibly having larger jet inclination angles. In a couple of cases, X-ray observations indicate the presence of relativistic beaming in contrary to the VLBI measurements made with the European VLBI Network (EVN). J1420$+$1205 is a prominent example, where our $30-100$\,mas-scale enhanced Multi Element Remotely Linked Interferometer Network (e-MERLIN) radio observations revealed a rich structure reminiscent of a small radio galaxy. It shows a bright hotspot which might be related to the denser interstellar medium around a young galaxy at an early cosmological epoch.}

\keywords{galaxies: high-redshift, galaxies: active, radio continuum: galaxies, galaxies: individual: J1420+1205}

\jnlcitation{\cname{%
\author{Gab\'anyi  K\'E}, 
\author{Frey S}, 
\author{An T}, 
\author{et al.}} (\cyear{2021}), 
\ctitle{A small radio galaxy at $z=4.026$}, \cjournal{Astron. Nachr.}, \cvol{}.}


\maketitle


\section{Introduction}\label{sec1}

High-redshift ($z\gtrsim 4$) AGN play an important role in the study of the early Universe and the evolution of AGN. The mere existence of powerful quasars at or above redshift $6-7$ already places constraints on the growth of supermassive black holes, accretion processes and the possible seed black holes \citep[e.g.,][and references therein]{TS_2021}. Roughly $10$\,\% of AGN are radio-loud \citep{rl_fraction}, these can be studied at pc-scale resolution even at high redshifts thanks to the VLBI radio observation technique. 

The radio-loud AGN whose jet points at a small angle to the line sight (the viewing angle is typically $\phi \lesssim 10^\circ$) are called blazars. In other words, blazars are radio-emitting AGN where relativistic effects significantly alter the observable features of the jets. For this to happen, the jet viewing angle has to be inside the beaming angle, thus $\phi_\mathrm{rad} \lesssim 1/\Gamma$, where $\Gamma$ is the bulk Lorentz factor of the jet.
The blazar nature of a radio-emitting AGN can be verified through fitting the spectral energy distributions (SEDs), or via VLBI imaging observations. In the latter case, the detection of a compact radio emitting feature with brightness temperature exceeding the equipartition limit \citep[$\sim 5 \times 10^{10}$\,K,][]{equipartition} indicates relativistic effects. Additionally, rapid variability, and high optical polarization could also be signs of the blazar nature.

Several $z>4$ objects have been identified, which were classified as blazar candidates via their SED-fitting \citep[for details see e.g.,][]{Swift_J1420}. However, VLBI observations failed to reveal a compact high-brightness temperature feature in two of them (J1420$+$1205, J2220$+$0025). They rather showed widely-separated double structures \citep{Cao2017}. Similar features were detected by \cite{Coppejans2016} in another $z>4$ source, J1548$+$3335, which was recently measured by {\it Chandra} to have the highest X-ray luminosity in a sample of $15$ radio-AGN at redshifts $4.5<z<5.0$ \citep{Snios2020}. We conducted radio interferometric observations of these three sources with e-MERLIN to map their intermediate-scale radio-emitting structures. Preliminary results of the observations were published  in \cite{EVN2018}. 

Here, we focus on J1420$+$1205 located at a redshift of $z=4.026$ \citep{sdss_dr14}\footnote{We note that a slightly different value, $z=4.034$, was reported previously in \cite{sdss_dr7} and used by us in \cite{EVN2018}.}. In the following, we use a $\Lambda$CDM cosmological model with $H_0=70 \mathrm{\,km\,s}^{-1} \mathrm{Mpc}^{-1}$, $\Omega_\mathrm{m}=0.27$ and $\Omega_\Lambda=0.73$. We define the radio spectral index $\alpha$ as $S\sim \nu^\alpha$, where $S$ is the flux density and $\nu$ is the observing frequency.

\section{Observations, data analysis and results}\label{sec2}

J1420$+$1205 was observed with e-MERLIN on 2017.07.15 and 2017.06.30 at $1.5$\,GHz, and on 2017.05.13 and 2017.06.27 at $5$\,GHz. For further details of the observations and data analysis, we refer to 
\cite{EVN2018}. 

\begin{figure*}[t]
\begin{minipage}{\columnwidth}
\centerline{\includegraphics[width=78mm, bb=0 2 440 515, clip=]{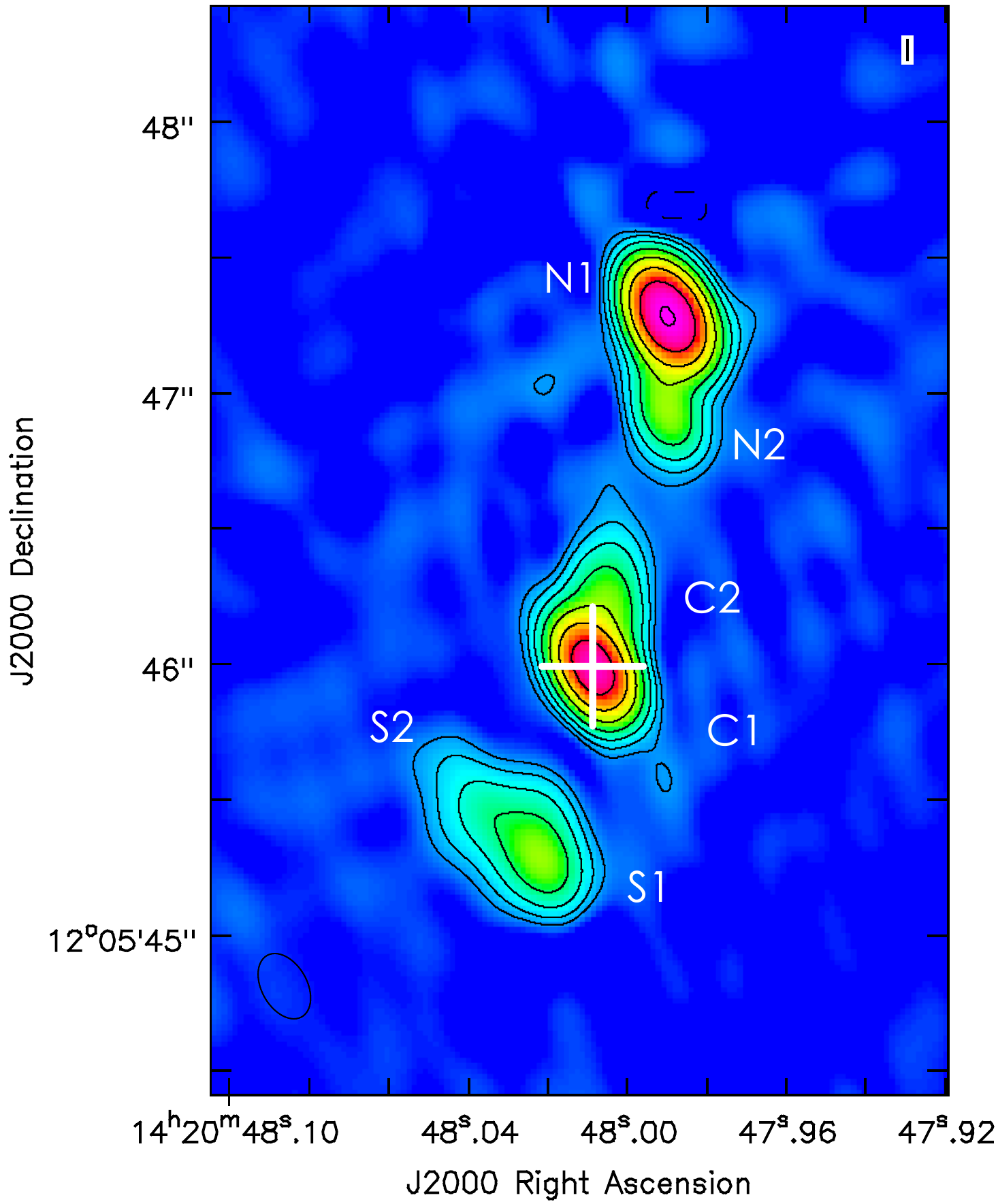}}
\end{minipage}
\hspace{-0.7cm}
\begin{minipage}{\columnwidth}
\centerline{\includegraphics[width=80mm, bb=0 2 370 425, clip=]{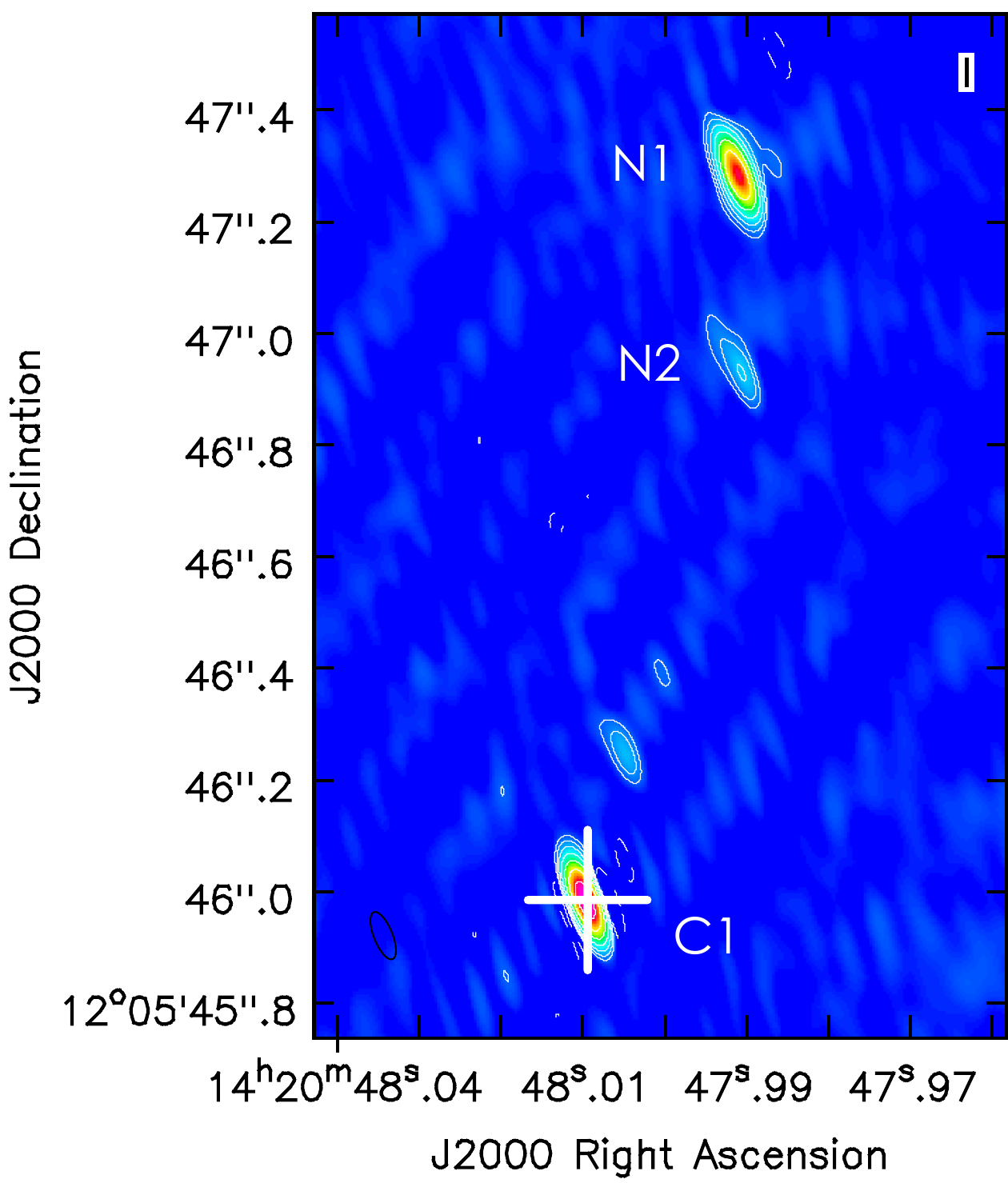}}
\end{minipage}
\caption{{\it Left}: $1.5$-GHz e-MERLIN image of J1420$+$1205. Peak brightness is $33.3 \mathrm{mJy\,beam}^{-1}$, the lowest positive contour level is drawn at $0.3 \mathrm{mJy\,beam}^{-1}$ corresponding to $5$ times the image noise level, further contour levels increase by a factor of $2$. The restoring beam is $0.41'' \times 0.14''$ at a major axis position angle of $\mathrm{PA}=28^\circ$. {\it Right}: $5$-GHz e-MERLIN image of J1420$+$1205 Peak brightness is$14.8 \mathrm{mJy\,beam}^{-1}$, the lowest positive contour level is drawn at $0.15 \mathrm{mJy\,beam}^{-1}$ corresponding to $6$ times the image noise level, further contour levels increase by a factor of $2$. The restoring beam is $0.09'' \times 0.03''$ at $\mathrm{PA}=22^\circ$. White crosses mark the {\it Gaia} EDR3 position \citep{Gaia_EDR3}. The restoring beams are shown in the lower left of each image. \label{eMERLIN}}
\vspace{-0.65cm}
\end{figure*}

At $1.5$\,GHz, we detected three main radio-emitting regions, designated as N, C, and S in Figure \ref{eMERLIN} left. The total extent of the radio source is $\sim2.5''$, corresponding to a projected linear size of $\sim 17.9$\,kpc in the redshift of the source. The brightest pixel is located in region N, while according to the {\it Gaia} EDR3 \citep{Gaia_EDR3,gaia_mission} optical coordinates (shown by white crosses at Figure \ref{eMERLIN}), the optical emission originates from region C. At $5$\,GHz, only the two regions N and C could be detected (see Figure \ref{eMERLIN} right). 

We used the {\bf {\sc difmap}} software \citep{difmap} to describe the brightness distribution by fitting the calibrated visibility data with circular Gaussian model components  in each spectral window independently. The $8$ spectral  windows of the $1.5$-GHz data can be consistently fitted with $6$ model components, using two components in each region. At $5$\,GHz, three Gaussian components were needed to describe the data of the two lower of the $4$ spectral windows (central frequencies $4.88$\,GHz and $5.01$\,GHz), two for region N and one for region C. These components are positionally coincident with N1, N2, and C1 identified at $1.5$\,GHz. To adequately fit the data of the two upper spectral windows (central frequencies $5.14$\,GHz and $5.26$\,GHz), $2$ components, N1 and C1 were sufficient. The component designations are depicted in Figure \ref{eMERLIN}. The errors of the component parameters were calculated according to \cite{error}.

We calculated the in-band spectral indices for the components which were detected at least in three spectral windows within one band. Additionally, we calculated the spectral indices for the positionally coincident components through the combined frequency range (Table \ref{specindex}).

\begin{center}
\begin{table}[t]%
\centering
\caption{The radio spectral index of the components of J1420$+$1205. \label{specindex}}%
\tabcolsep=0pt%
\begin{tabular*}{20pc}{@{\extracolsep\fill}lccc@{\extracolsep\fill}}
\toprule
\textbf{ID} & \textbf{$\alpha_{1.29}^{1.73}$}  & \textbf{$\alpha_{4.88}^{5.26}$}  & \textbf{$\alpha_{1.29}^{5.26}$} \\
\midrule
C1 & $-0.44\pm0.04$ & $-0.69\pm0.35$ & $-0.48\pm0.01$ \\
C2 & $+0.10\pm0.21$ &  -- & --  \\
N1 & $-0.97 \pm 0.05$  & $-0.55\pm0.42$  & $-1.01\pm0.01$  \\
N2 & $-1.94\pm0.14$  & --\tnote{$\dagger$} & $-2.00 \pm 0.06$-\tnote{$\ddagger$}  \\
S1 & $-2.90\pm0.25$  & --  & -- \\
S2 & $-1.46\pm0.50$ & --  & --  \\
\bottomrule
\end{tabular*}
\begin{tablenotes}
\item[$\dagger$] N2 could only be fitted in two spectral windows of the $4.9$-GHz band, thus the spectral index is not well constrained.
\item[$\ddagger$] N2 could not be fitted at the two upper spectral windows, this spectral index is calculated between $1.29$\,GHz and $5$\,GHz.
\end{tablenotes}
\vspace{-0.60cm}
\end{table}
\end{center}

\section{Discussion and Summary}\label{sec3}

J1420$+$1205 has been detected in the Faint Images of the Radio Sky at Twenty-Centimeters survey \citep[FIRST,][]{FIRST_cat} with a flux density of $(87.3\pm 6.6)$\,mJy as an unresolved object. In our e-MERLIN observation, the sum of the flux densities of the fitted model components in the spectral window closest to the FIRST frequency (at $1.414$\,GHz), is $(80.8\pm0.4)$\,mJy, agreeing within the errors with the FIRST value. At $\sim 4$\,mas resolution, $1.7$-GHz EVN observations in 2015 Oct detected two features corresponding to C1 and N1 with a total flux density of $47.9$\,mJy \citep{Cao2017}. The total flux density detected by the e-MERLIN at the closest frequency, $1.67$\,GHz is $(69.6\pm 0.4)$\,mJy. The difference is mainly due to the additional features (C2, N2, S1, and S2) revealed by the lower resolution observation, while the flux densities of C1 and N1 agree roughly within the errors.

At $5$\,GHz, only component C1 with a brightness temperature of $T_\textrm{b}\sim1.6\times10^9$\,K is detected with the EVN, hinting the steeper spectral index and/or possibly more extended nature of component N1. Our e-MERLIN spectral indices confirm the former, across the two bands N1 has a steep spectrum, with $\alpha_\mathrm{N1}=-1.01\pm0.01$, while C1 has a flatter spectrum with $\alpha_\mathrm{C1}=-0.48\pm0.01$ (Table \ref{specindex}). The flux density of C1 in the 5-GHz EVN observation is significantly lower, $\sim 8$\,mJy compared to the one measured by e-MERLIN, $(14.4\pm0.2)$\,mJy. In principle, this can be due to resolution effect and/or source variability. The lack of variability seen at $1.7$\,GHz makes the latter rather unlikely. It is more probable that, similarly to N1, the VLBI observation resolved out significant flux density which, however, could be recovered by e-MERLIN. 

The {\it Gaia} optical point source coincides with C1 (and its coordinates agree with the more precise EVN localization of the compact feature), thus it is most probably the position of the AGN core, and the radio emission is related to the jet base. Regions N and S probably correspond to the lobe emissions. The reason for the much brighter northern arm may arise due to asymmetric structure in the interstellar medium of the host galaxy or due to light-travel time effect \citep{flux_ratio,arm-length}. In the latter case, the advancing, thus brighter lobe is seen farther away from the central region in the sky, thus the arm-length ratio of the brighter to fainter lobe exceeds unity. According to this picture, region N is on the advancing side of J1420$+$1205.

\begin{figure}[t]
	\centerline{\includegraphics[height=78mm, bb=40 50 550 730, angle=-90, clip= ]{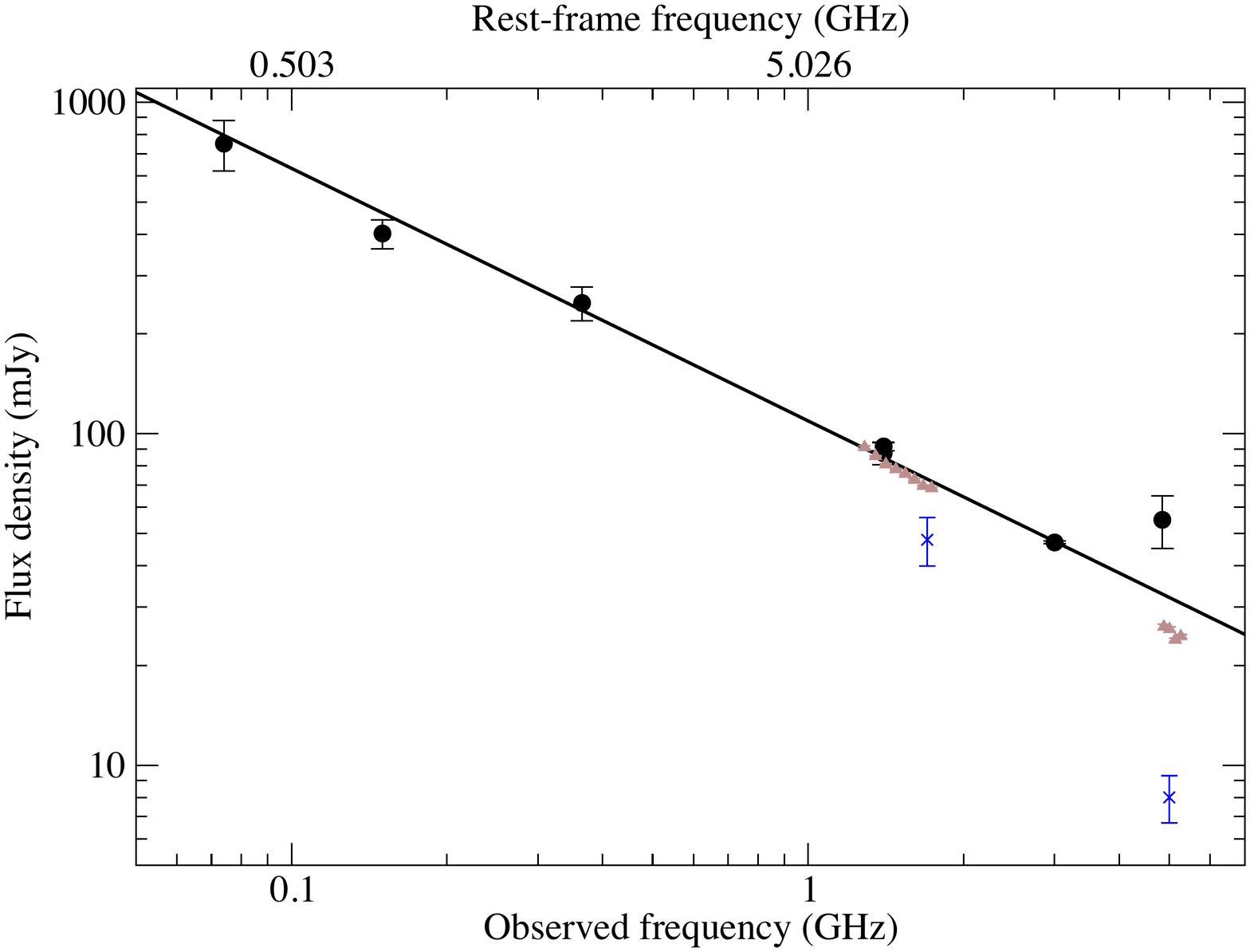}}
	\caption{Broad-band radio spectrum of J1420$+$1205, with a fitted power law (solid line) to the archival flux density measurements (black dots). These are the followings: VLA Low Frequency Sky Survey \citep{vlss}, TIFR GMRT Sky Survey-ADR \citep{gmrt}, the Texas Survey \citep{texas}, the NRAO VLA Sky Survey \citep{nvss}, FIRST \citep{FIRST_cat}, preliminary data from the VLA Sky Survey \citep{vlass}, and the 87GB catalog \citep{gb6, newcat}. For comparison, the e-MERLIN (brown triangles) and EVN \citep[blue crosses,][]{Cao2017} total flux densities are also shown. \label{broad}}
	\vspace{-0.70cm}
\end{figure}

We calculated the $1.4$-GHz radio power of J1420$+$1205 from the flux density measured at $1.414$ GHz with e-MERLIN, 
assuming a spectral index of $-0.76\pm0.04$ based on the radio flux density data collected from the literature, covering a wide frequency range (Figure \ref{broad}). The obtained value, $9.2 \times 10^{27} \mathrm{\,W\,Hz}^{-1}$ together with the linear extent, $\sim 17.9$\,kpc measured in the e-MERLIN image indicate that the source is a high-power compact steep-spectrum source and can be regarded as a young (and smaller) version of a Fanaroff--Riley type 2 radio galaxy according to \cite{AnBaan}.


Based upon X-ray and UV measurements obtained with the {\it Swift} satellite, \cite{Swift_J1420} concluded that the SED of J1420$+$1205 can be best described as of a beamed blazar source. \cite{Chandra_obs} analysed the multi-wavelength properties of $17$ $z>4$ radio-loud AGN, including J1420$+$1205, for which {\it Chandra} snapshot observations were obtained. They found an X-ray emission enhancement compared to similar radio-loud quasars at lower redshifts. They showed that this cannot be explained by the inverse Compton upscattering of the Cosmic Microwave Background photons (as that would produce even more X-ray enhancement).

Our radio interferometric observations showed that J1420$+$1205 is not a blazar but more likely a young radio galaxy. This is in tension with previous X-ray measurements of the source where the X-ray emission was explained as being related to the relativistically beamed jet. The X-ray emission may originate either from a region where the jet viewing angle is much lower than that was deduced from our radio observations, or (partly) created through the interaction between the jet and the surrounding medium in the lobes. However, in the latter case no bright radio emission is expected \citep{Ghisellini_2014}. In the future, higher resolution X-ray observation better constraining the location of the emission may discriminate between these scenarios.



\section*{Acknowledgments}
The e-MERLIN is a National Facility operated by the University of Manchester at Jodrell Bank Observatory on behalf of STFC. Scientific results presented in this publication are derived from the e-MERLIN project code: CY5211. This work has received funding from the European Union's Horizon 2020 research and innovation programme under grant agreement No 730562 (RadioNet), and the NKFIH (OTKA K134213 and 2018-2.1.14-T\'ET-CN-2018-00001). We thank for the usage of the ELKH Cloud (previously known as MTA Cloud).







\bibliography{ref_highz}

\section*{Author Biography}
\begin{biography}{}{\textbf{K. \'E. Gab\'anyi.} Adjunct lecturer at the ELTE University (Budapest, Hungary). Her main research interests are radio interferometric studies of dual or recoiling and high-redshift AGN.}
\end{biography}

\end{document}